\documentclass{PoS}

\title{The impact of stellar rotation on the CNO abundance patterns in the Milky Way
at low metallicities}

\ShortTitle{The impact of stellar rotation on the MW CNO abundance patterns 
at low metallicities}

\author{\speaker{Cristina Chiappini}\thanks{Currently on sabbatical at
        the Observatoire de Geneve, Sauverny, Switzerland}\\
        Osservatorio Astronomico di Trieste, Via G. B. Tiepolo 11, I - 34131 Trieste, Italia\\
        E-mail: \email{Christina.Chiappini@obs.unige.ch}}

\author{Hirschi, R.\\
         Dept. of Physics and Astronomy, University of Basel, CH-4056, 
         Basel, Switzerland\\}

\author{Matteucci, F.\\
        Dipartimento di Astronomia, Universit\'a degli Studi di Trieste, 
Via G. B. Tiepolo 11, I - 34131 Trieste, Italia\\}

\author{Meynet, G., Ekstr\"om, S., Maeder, A.\\
        Observatoire Astronomique de l'Universit\'e de Gen\`eve, CH-1290, Sauverny, Switzerland\\}

\abstract{
We investigate the effect of new stellar models, which take 
  rotation into account, computed for very low metallicities
  (Z = 10$^{-8}$) on the chemical evolution
  of the earliest phases of the Milky Way. We check the impact of
  these
new stellar yields on a model for the halo of the Milky Way that can
reproduce the observed halo metallicity distribution. In this way we
try to better constrain the ISM enrichment timescale, which
was not done in our previous work (\cite{Chia06}).
The stellar models adopted in this work were computed
  under the assumption that the ratio of the initial rotation velocity
  to the critical velocity of stars is
  roughly constant with metallicity. This naturally leads to faster 
rotation at lower metallicity, as metal poor stars are more compact than metal
  rich ones. We find that the new Z = 10$^{-8}$
  stellar yields computed for large rotational velocities 
have a tremendous impact on the interstellar medium
  nitrogen enrichment for log(O/H)+12 $<$ 7 (or [Fe/H]$<$ $-$3). 
We show that upon the inclusion of the new stellar calculations
in a chemical evolution model for the galactic halo with infall and outflow, 
both high N/O and C/O ratios
  are obtained in the very-metal poor metallicity range in agreement
  with observations. Our results
  give further support to the idea that stars at very low
  metallicities could have initial rotational velocities of the order
  of 600-800\,km\,s$^{-1}$. An important contribution to N from AGB
  stars is still needed in order to explain the observations at
  intermediate metallicities. One possibility is that AGB stars at
  very low metallicities also rotate fast. This could be tested in the
  future, once 
  stellar evolution models for fast rotating AGB stars will be
  available.}

\FullConference{International Symposium on Nuclear Astrophysics --- Nuclei in the Cosmos --- IX\\
		 June 25-30 2006\\
		 CERN, Geneva, Switzerland}

\begin{document}

\section{Introduction}
Recent measurements of nitrogen abundances in metal-poor stars by
Spite et al.
\cite{S05}
show a large N/O ratio suggesting high levels of
production of primary nitrogen in massive stars. 
Moreover, the N/O abundance ratios in metal-poor
stars exhibit a large scatter (roughly 1 dex, much larger than
their quoted error bars) although none of the stars measured so 
far has N/O ratios as low as the ones observed in DLAs.

Last year, Chiappini et al. \cite{Chia05} studied the
implications of this new data set for our
understanding of the nitrogen enrichment in the Milky Way (MW). 
By the time the
latter paper was published there was no set of stellar yields able to 
explain the very metal-poor data of \cite{S05}. Using the so-called Z=0
population III stellar yields available in the literature (which in
some cases predict large N yields for non-rotating massive stars 
with masses around 20 M$_{\odot}$ - e.g. \cite{Chie04}) did not solve the problem
either. The main reason is that in population III non rotating models  
the N production is confined to a 
very narrow mass range and hence chemical evolution models computed
with such prescriptions predict a peak in the [N/Fe] ratios at very
low metallicities
followed by a strong decrease already before 
[Fe/H] $\sim -$3 in disagreement with the observations of \cite{S05} (see \cite{Ba06}). 
This happens even
when allowing the Z=0 contribution to be valid up to a threshold (Z$_{tr}$)
metallicity above which population III stars stop forming of the order 
of Z$_{tr}$=10$^{-6}$. This Z$_{tr}$ metallicity is probably already
too high as the physics adopted in the Z=0
models would be valid only up to metallicities 
Z$\sim$10$^{-10}$, which represents 
the metallicity below which massive stars first enter the phase of 
H-burning via the pp chain, followed by the 3$\alpha$ reaction, which
then allows the CNO cycle to proceed, as in Z=0 (population III) stars.

Given the above discussion, \cite{Chia05} suggested 
that the most promising way to account for
the new data was to assume that stars at low metallicity rotate sufficiently
fast to enable massive stars to contribute much larger 
amounts of nitrogen. It was shown that even when adopting the
Meynet \& Maeder \cite{M02} stellar yields (these authors had computed
stellar yields down to Z=10$^{-5}$) where the nitrogen production 
is increased in stars of all masses due to rotation, it was not
possible
to explain the observations of \cite{S05}. The approach of \cite{Chia05} was to
assume that the stellar yields computed by \cite{M02} were
valid down to Z=10$^{-5}$, but that below this metallicity some 
mechanism should be able to increase even more the N yields. 
It was then predicted that massive stars born with
metallicities below Z=10$^{-5}$ should produce a factor between 10 and
a few times 10$^2$ more nitrogen (depending on the stellar mass) than the values 
given by \cite{M02} for Z=10$^{-5}$ and
$\upsilon^{\rm ini}_{\rm rot}=300$\,km\,s$^{-1}$ in order to reproduce
the {\it mean} locus of the Spite et al. data in a log(N/O) vs. log(O/H)+12 diagram.

The physical motivation for the approach described above
would be an increase of the rotational velocity in very metal-poor
stars \cite{M99, M06} and hence an increase in the nitrogen
yields. 
In this framework it is possible to understand the apparent
 contradictory
finding of \cite{S05} namely, a large scatter in N/O and the
 almost complete lack of scatter in [$\alpha$/Fe] ratios of the same
 very metal-poor halo stars \cite{Ca04}.
As suggested in \cite{Chia05}, if the nitrogen production in very metal-poor
massive stars depends strongly on the rotational velocity of the star,
this could explain the scatter in the N/O abundance ratios: the
scatter would reflect the distribution of the stellar rotational
velocities as a function of metallicity whereas the [$\alpha$/Fe]
ratios would remain unchanged.

Whether the above suggestions were physically plausible remained to 
be assessed by stellar evolution models computed at lower
metallicities, which took rotation and mass loss into account.
Novel stellar evolution models have been 
computed for metallicities Z=10$^{-8}$ \cite{Hi06} for massive stars. 
The new calculations show that if the stars at all Z start their
life on the zero age main sequence (ZAMS) with on the average a
fraction of $\sim$\,0.5 the
critical velocity, the low Z stars easily reach break-up velocity
during MS evolution (see also Meynet et al., this conference). 
Fast rotation also contributes to produce a more efficient
mixing at lower Z, which leads to a large production of N in massive
stars. We note that the new models with fast rotation also predict 
different type of supernova progenitors,
final remnants and yields in heavy elements at very low Z (\cite{M06}
and references therein). 

In a recent paper, Chiappini et al. \cite{Chia06} computed 
chemical evolution models adopting these new stellar prescriptions
and show that these new models can explain not only the high N/O in
very metal poor stars but also predict an increase in C/O for
decreasing metallicity.  However the model computed by \cite{Chia06} did not include outflows
during the halo phase thus leading to an increase of [Fe/H] with time
faster than what would be obtained with an outflow model. Moreover, as
shown by \cite{P03}, halo models that can reproduce the
metallicity distribution of halo stars should include both inflow and outflow.

Here we compute such a model and compare our new results with the ones
we obtained in \cite{Chia06}. Finally, in \cite{Chia06}
we considered only the so-called
{\it normal} very metal-poor stars, since our goal was to explain the
mean ISM enrichment at very low metallicities. Here we place the
carbon rich ultra metal-poor stars \cite{B05} into
this context and discuss the implications for the CNO nucleosynthesis.

\section{Results obtained with a "halo" model without outflow}

In this section we show the results obtained in \cite{Chia06} who computed
the evolution of the CNO elements, in the
solar vicinity, predicted by a chemical evolution model computed with different 
stellar yield sets for metallicities below Z=10$^{-5}$ for massive
stars\footnote{For low and intermediate mass
stars we adopt \cite{M02} stellar yields with rotation, and assume that their table
computed for Z=10$^{-5}$ is valid down to Z=0., as in Chiappini et al. 2003.}. The 
adopted stellar yields for Z $\leq$ 10$^{-5}$ are shown in Fig. 1.

In Fig. 1, filled
squares show stellar yields  of \cite{M02} for their lowest
metallicity case (Z=10$^{-5}$) resulting from models with rotation
($\upsilon^{\rm ini}_{\rm rot}=300$\,km\,s$^{-1}$),
while open symbols stand for models computed with $\upsilon^{\rm ini}_{\rm rot}=0$\,km\,s$^{-1}$. The asterisks connected by the long-dashed line show
the {\it ad hoc} stellar yields for metallicities Z$<$10$^{-5}$
adopted in the {\it heuristic model} of \cite{Chia05} for massive stars. 
The dots show the
new results obtained by \cite{Hi06}
for Z=10$^{-8}$ for massive stars.
The stellar yields for the Z=10$^{-8}$
case were computed according to the following assumption: {\it stars begin
    their evolution on the ZAMS with approximately the same angular
momentum content, regardless of their metallicity}. At solar metallicity,
    observations indicate a mean rotational velocity of a 60
    M$_{\odot}$ star on the MS of the order of 200\,km\,s$^{-1}$, which corresponds to an initial angular
momentum of the order of 2 $\times$ 10$^{53}$\,g\,cm$^2$\,s$^{-1}$ (see
   \cite{M06} for details). 
At a metallicity of Z=10$^{-8}$ this corresponds to
$\upsilon^{\rm ini}_{\rm rot}=$ 800\,km\,s$^{-1}$. In other words, we adopt a rotational
velocity such that
    the ratio between $\upsilon^{\rm ini}_{\rm rot}/\upsilon_{\rm breakup}$ remains almost constant
    (around 0.5) with mass and metallicity (see \cite{Hi06} for details).
The very interesting result is that the new computations by Hirschi \cite{Hi06} for Z=10$^{-8}$ predict a large increase of the N
    yields for stars above 20 M$_{\odot}$, similar to the {\it ad hoc}
    yields
of \cite{Chia05}.

\begin{figure}
\centering
\includegraphics[width=6cm,angle=0]{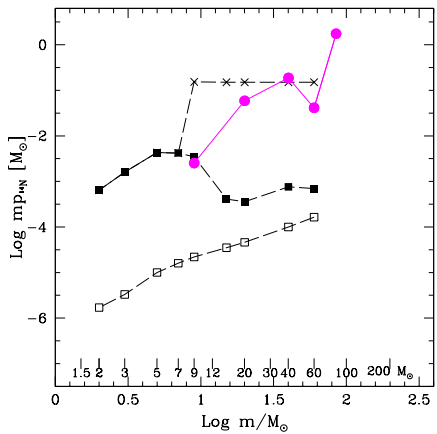}
\caption{
Stellar yields for $^{14}$N at low metallicities for the
  whole stellar mass range. The yields of \cite{M02} for stellar models with
  and without rotation for Z = 10$^{-5}$ are shown by filled and open
  squares respectively. The asterisks connected by the long-dashed
  line show the {\it ad hoc} stellar yields adopted for Z$<$ 10$^{-5}$ in the
  heuristic model of \cite{Chia05} (see text). The filled circles show the new stellar
  yields computed for massive stars born at Z=10$^{-8}$. 
The new computations of Hirschi \cite{Hi06} for Z=10$^{-8}$ lead to 
a large increase of N yields for masses above $\sim$20 M$_{\odot}$,
  similar to the predictions of \cite{Chia05}. This agreement is striking
  considering that they were obtained from completely different approaches.}
\end{figure}

Figure 2 shows the predicted evolution of our chemical evolution model
without outflow (see model details in \cite{Chia06}) for N/O and C/O  for 3 different
cases: a) solid line (black) - a model computed under the assumption that the
lowest metallicity yields table computed by \cite{M02} for Z=10$^{-5}$ and
$\upsilon^{\rm ini}_{\rm rot}=300$\,km\,s$^{-1}$ is valid down to Z=0.
In this case the model flattens (upper panel) for log(O/H)+12 $<$ 6.6
due to the contribution by massive stars to the nitrogen predicted by
stellar models where rotation is included. However, it is not enough
nitrogen to explain the observations. This model also predicts a C/O
ratio that decreases with decreasing metallicity; b) dot-dashed line
(red - upper pannel) - the {\it heuristic} model of \cite{Chia05} were an {\it ad
  hoc}
larger yield of nitrogen is assumed for metallicities below
Z=10$^{-5}$ (as shown by the asterisks in Fig.1); c) dashed curve (black) -
the model of \cite{Chia06} computed with 4 yield tables for
Z = 0.020, 0.004 and 10$^{-5}$ from \cite{M02} (for $\upsilon^{\rm ini}_{\rm
  rot}=300$\,km\,s$^{-1}$) and the new stellar
yields of \cite{Hi06} for Z=10$^{-8}$ (for $\upsilon^{\rm ini}_{\rm
  rot}=500-800$\,km\,s$^{-1}$ depending on the stellar mass). The
latter model is able to explain the \cite{S05} data and confirms the
suggestion made in \cite{Chia05} that rotation could be the solution for the
{\it N-problem}. For the C/O ratio this model predicts an increasing 
C/O ratio with decreasing O/H.

\begin{figure}
\centering
\includegraphics[width=8cm,angle=0]{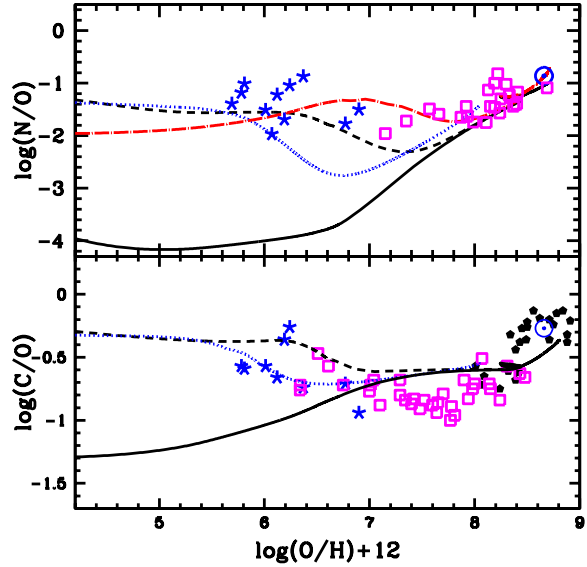}
\caption{Solar vicinity diagram log(N/O)
  vs. log(O/H)+12 (upper pannel) and log(C/O) (lower panel). 
The data points are from: \cite{I04} for N/O and \cite{Ak04} for C/O (large
  squares), \cite{S05} for N/O and \cite{Ca04} for C/O (asterisks) and from \cite{N03} (filled pentagons). Solar
abundances (\cite{A05} and references
therein) are also shown. For the meaning of the different curves, see
Section 2.}
\end{figure}

Figure 3 shows the evolution of [C/Fe], [N/Fe] and [O/Fe] as functions of
[Fe/H] for the same models (for the meaning of the dotted (blue) curve 
both in Figs. 2 and 3, see next section).

\begin{figure}
\centering
\includegraphics[width=8cm,angle=0]{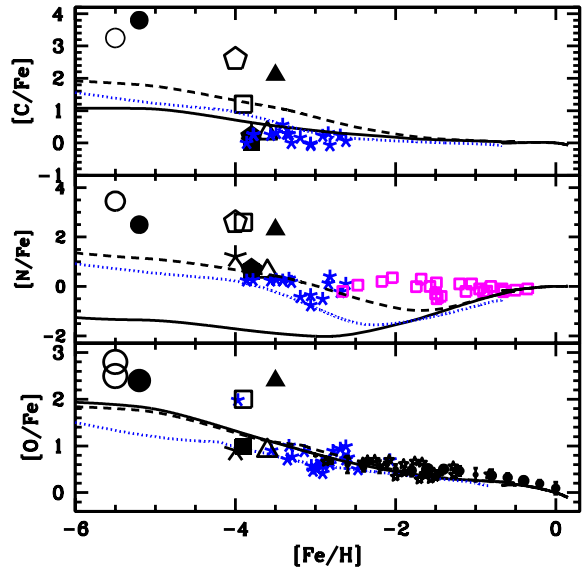}
\caption{The evolution of [C/Fe], [N/Fe] and [O/Fe] as functions of
  [Fe/H]. For the meaning of the curves, see Section 2. The small asterisks
  show the data of Spite et al. \cite{S05}. In the middle panel, the small squares
  are data from \cite{I04}. In the bottom panel the small dots are from \cite{Mel02}. 
The large symbols show some of the very metal poor stars discussed in Beers and 
Christlieb review (\cite{B05} and references therein). These stars are labelled as
  follows:G77 61 (open pentagon), CS22885-096 (filled pentagon), CS22949-037
  (open square), BS 16467-062 (asterisk), CS22172-002 (filled square),
  CS22968-014 (open triangle) and CS2 29498-043 (filled triangle). }
\end{figure}

The results shown both, in Fig. 2 and Fig. 3 for the dashed curve
implies that faster
rotation is able to explain the observations of Spite et al. (\cite{S05}  - asterisks
in these figures), accounting for
the almost solar N/O ratios found by these authors for metal-poor {\it
  normal} halo stars. An interesting result is also seen for C/O,
where the new model leads to an upturn of this abundance ratio at low
metallicities that can be explained in two ways: a) the high
production of primary N implies a very active H-burning shell, which
contributes a large part of the total luminosity of the star. As a
consequence, part of the total luminosity compensated by the energy
produced in the helium core is reduced, making the average core
temperature and thus the efficiency of the
$^{12}$C($\alpha,\gamma$)$^{16}$O reaction lower; b) more efficient
mixing also leads to larger mass loss, decreasing the He-core
size. Higher C/O ratios are thus obtained at the end of the He-burning phase.
 
\section{Results obtained with a halo model with outflow}

The chemical evolution model discussed above nicely fits the observed
data for N/O and C/O ratios at low metallicities. 
However, as nicely shown by \cite{P03}, the enrichment timescale
(i.e. the time that it takes for the ISM to reach a certain
metallicity
or, equivalently, the lower mass contributing to the ISM enrichment
at a given log(O/H)+12 or [Fe/H]) is still unknown in the case of
the galactic halo. The situation is different for the solar vicinity
where we have many more observational constraints. Processes like
outflows
can change the enrichment timescale. However, in the last years more
data on metal poor halo stars have been obtained (see Beers, this
conference) and it is now possible to better estimate the halo
metallicity distribution. The latter is an important constraint on the
enrichment timescale of the galactic halo.

\begin{figure}
\centering
\includegraphics[width=10cm,angle=0]{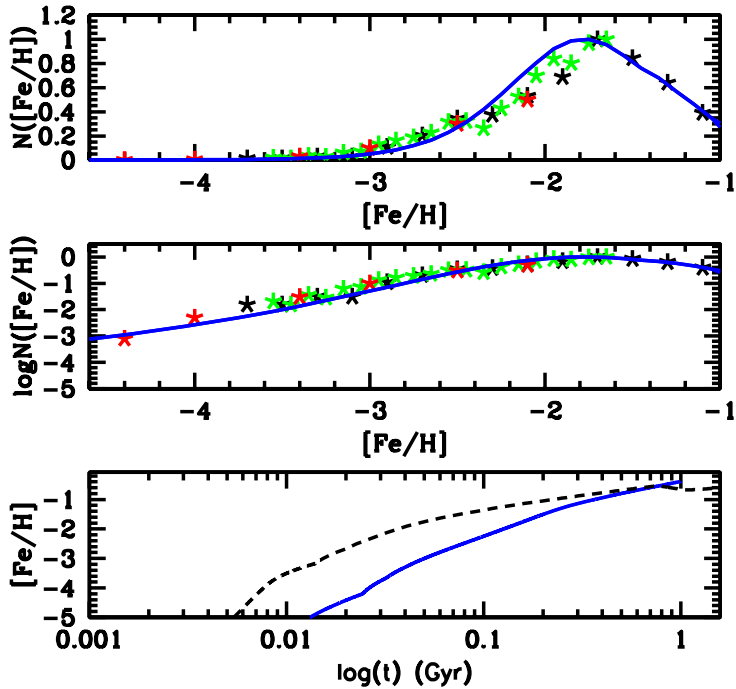}
\caption{Upper panel: Halo metallicity distribution normalized to one
  (linear). Middle panel: the same as in the upper panel, but in
  log-scale. The data points are from  \cite{RN91} (black), 
\cite{B05} and Beers and Christlieb (priv. com) (green)
 (upper pannel) and from \cite{N99} (red). 
The data points of Beers and collaborators were
shifted by $-$0.4 dex to match the distribution peak of
 \cite{RN91} around [Fe/H]=$-$2. 
Lower panel: the evolution of the metallicity with time
  (for the meaning of the curves, see text).}
\end{figure}

In this section we show the results obtained with a different model
for the galactic halo. In this case we assume a) a gaussian infall 
($f(t) \propto e^{(t-t_0)^2/2 \sigma^2}$, with $t_0 = $0.1 Gyr and $\sigma
=$ 0.05); b) an outflow rate of 8 times the star formation rate; c) a
Schmidt law for the star formation rate. The halo metallicity
distribution obtained with the latter model is shown in Fig. 4 and it is in good agreement
with observations. In the lower panel of Fig. 4
it is seen that this model (solid blue line) predicts a slower
ISM enrichment compared to the model without outflow, as expected.
While in our previous model a 8 M$_{\odot}$ star would die at a metallicity
[Fe/H]$\sim -$2.2, in the new model with outflow the same star would
die when the metallicity in the ISM was [Fe/H]$\sim -$3.8. The
resulting predictions for the abundance ratios of C/O, N/O, C/Fe, N/Fe and O/Fe in this case are shown
in Figs. 2 and 3 by the dotted (blue) line.

As it can be seen in these figures, this model gives the same
qualitative results as the ones of our previous models (\cite{Chia06} and previous
Section). This means 
that even if the AGB start contributing earlier on (in terms
of metallicity) they are still unable to contribute an important
amount of N at the very low metallicities sampled by the Spite et
al. data. Quantitatively, things change a little bit and the 
lack of agreement both in the N/O and N/Fe plots at intermediate 
metallicities (already present in our earlier models, see \cite{Chia06}) 
becomes even more evident. This means that although these models can
well explain the quantities of N at very low metallicities, they
underproduce this same element at intermediate metallicities.

However, as already stressed in \cite{Chia06}, here we did not take
fully into account the contribution of the AGBs to the N enrichment.
First, for the low and intermediate mass stars we are adopting the stellar
yields of \cite{M02} where no 3rd dredge-up was included, and thus
these yields represent
a lower limit for the N yields. Second, AGB stars could also rotate
faster at lower metallicities and this was not implemented in our
models as stellar models for fast rotating AGBs in very metal poor
environments are still not available. Third, we also did not include
the contribution of the super-AGB stars which supposedly contribute
essentially to N. In the same intermediate metallicity
range we see that our models overestimate the C/O ratios. There is
thus room to increase N through processes that consume C as it is for instance the
case of HBB.

Finally,  we notice that the so called "carbon rich very metal poor stars" \cite{B05} are still
located above our chemical evolution model curves (see
Fig.3). However, for N, their locus is much closer to what is expected
for the ISM evolution at such very low metallicities once fast
rotation is taken into account than in the case of slowly rotating
models (solid line). This means that the "extra" N needed to explain
stars like the two most metal poor ones is less than usually
assumed. As shown by \cite{M06} and \cite{Hi06} it is probably
possible to account for this difference by invoking metal enrichment
only by stellar winds for stars with M$\geq$30M$_{\odot}$.

\section{Conclusions}

In this work, we computed chemical evolution models adopting the 
very recent calculations of Hirschi \cite{Hi06}
for the evolution of massive stars at very low metallicities
under
the assumption of an almost constant ratio $\upsilon^{\rm ini}_{\rm rot}/\upsilon_{\rm breakup}$ as a
function of metallicity (i.e. where the $\upsilon^{\rm ini}_{\rm rot}$ increases towards lower
metallicities). In such a framework,
massive stars can produce large amounts of nitrogen.
 
At present, this is the only way to
explain the high nitrogen abundances
 measured recently in {\it normal} halo stars.
This gives further support to the idea that 
stars rotate faster at very low metallicities.
The new stellar evolution models
also produce some extra carbon at Z $\leq$ 10$^{-5}$. As a
consequence, an upturn is produced in the C/O at low metallicities.
This result is obtained without the need of introducing
population III stars (here understood as Z=0 stars with a flatter IMF). 

The only alternative to explain the observations showing a high N abundances
 in very metal poor normal stars\footnote{Here, "normal" stands for the assumption that these stars show a pristine N abundance which reflects the one of  the ISM gas from which
 these stars formed and did not come from any internal mixing process or binary mass transfer.}
 would be if AGB stars would contribute to the ISM enrichment before
 [Fe/H]$\sim -$4. This would mean that  the timescales for the
 chemical evolution of the halo would be very different from the
 ones of \cite{Chia06}.  Models assuming outflows could delay the ISM enrichment so that AGBs
 would have time to contribute. Here we show that a model with outflow,
which is able to fit the observed halo metallicity distribution and in which the ISM is enriched on longer timescales than the ones computed by \cite{Chia06}, still
gives the same qualitative result. In fact, this model still requires to invoke fast rotation at very low metallicities  to explain the high N/O observed in metal poor halo stars. Future surveys of very metal poor stars (see Beers this conference) will certainly lead to a more complete halo metallicity
distribution and will be extremely useful to constrain even better the halo enrichment timescale.

Here we show the results for the mean ISM enrichment history, assuming that at a given
metallicity all stars of a given mass rotate at the same velocity.
Further computations are in progress, and we envision
to compute chemical evolution models in which a
distribution of rotational velocities is assumed for each metallicity. These models
would predict a scatter in N/O to be compared to the observed one.

\begin{acknowledgments}
C. C. acknowleges partial support from INAF PRIN grant CRA 1.06.08.02
\end{acknowledgments}

\end{document}